\documentclass{ifacconf}

\usepackage{amssymb}
\usepackage{graphicx}      
\usepackage{natbib}        
\usepackage{amsmath}
\usepackage{pgfplots,tikz}
\usepackage{mathtools}

\usepackage{tikz}
\usetikzlibrary{calc}
\usetikzlibrary{shapes}

\usepackage{url}
\usepackage{algorithm}
\usepackage{algpseudocode} 
    
\newcommand{\mc}{\mathcal}
\newcommand{\ol}{\overline}
\newcommand{\ul}{\underline}

\newcommand{\defineas}{\coloneqq}

\newtheorem{remark}{Remark}
\newtheorem{definition}{Definition}

\usepackage{graphicx}      
\usepackage{natbib}        
\begin{document}
\begin{frontmatter}

\title{Dynamic Routing in Stochastic Urban Air Mobility Networks: A Markov Decision Process Approach
} 


\thanks[footnoteinfo]{This work was partially supported by the NASA University Leadership Initiative (ULI) under grant number 80NSSC21M071.}

\author[First]{Qinshuang Wei} 
\author[Second]{Yue Yu} 
\author[Third]{Ufuk Topcu}

\address[First]{The University of Texas at Austin, 
   Austin, TX 78712 USA (e-mail: qinshuang.wei@austin.utexas.edu.)}
\address[Second]{The University of Texas at Austin, 
  Austin, TX 78712 USA (e-mail: yueyu@utexas.edu)}
\address[Third]{The University of Texas at Austin, 
  Austin, TX 78712 USA (e-mail: utopcu@utexas.edu)}

\begin{abstract}                
Urban air mobility (UAM) is an emerging concept in short-range aviation transportation, where the aircraft will take off, land, and charge their batteries at a set of vertistops, and travel only through a set of flight corridors connecting these vertistops. We study the problem of routing an electric aircraft from its origin vertistop to its destination vertistop with the minimal expected total travel time. We first introduce a UAM network model that accounts for the limited battery capacity of aircraft, stochastic travel times of flight corridors, stochastic queueing delays, and a limited number of battery-charging stations at vertistops. Based on this model, we provide a sufficient condition for the existence of a routing strategy that avoids battery exhaustion. Furthermore, we show how to compute such a strategy by computing the optimal policy in a Markov decision process, a mathematical framework for decision-making in a stochastic dynamic environment. We illustrate our results using a case study with 29 vertistops and 137 flight corridors.


\end{abstract}


\end{frontmatter}

\section{Introduction}
Urban air mobility (UAM), an air transportation concept used for short trips within urban and suburban areas, is emerging as an important component in future transportation networks. 
The aircraft in UAM operate in a transportation network---named \emph{UAM network}---that consists of \emph{vertistops} (or \emph{vertiports}) installed on roofs of existing buildings with limited capacity~\cite{2018landscape, balakrishnan2018blueprint} and flight corridors connecting the vertistops~\cite{inrix, balakrishnan2018blueprint}.

A key challenge in routing UAM aircraft is the limited battery capacity of electric-powered aircraft. Unlike traditional gasoline-powered aircraft, electric-powered aircraft are preferable in UAM because of lower emissions and decreased fuel consumption \cite{goyal2018urban,7098414}. However, aircraft routing in UAM requires safe routing strategies that choose which flight corridors to use and which intermediate vertistops to land and charge the battery during a flight such that the battery is never exhausted, while adaptive to the existence of waiting queues and limited landing capacities at vertistops.    

Although existing works of ground electric vehicles (EVs) have investigated routing problems with battery considerations, there are several limitations to applying these studies to electric-aircraft routing problems due to the difference between ground and air vehicle operations. 
Prior works have studied routing services that search and reserve a charging station in~\cite{bessler2012routing}, routing with battery swapping strategies in~\cite{adler2014online, yang2015battery} with consideration of limited available batteries, routing problem that minimizes travel time cost and energy cost in~\cite{lin2016electric}, a joint EV routing and charging strategy that maximizes the revenue of EV users in~\cite{trivino2019joint}, and a method for computing the optimal route and charging schemes while considering interactions of power distribution and traffic network in~\cite{liu2022collaborative}. However, ground EVs do not worry about the battery consumption for waiting for the battery to be charged (or to be swapped) as in the aerial case, while the aircraft needs to consume power when hovering and waiting in the air. 
Moreover, they also ignore the charging time cost. Although a number of studies of ground EVs prefer to swap the battery, it is preferable to recharge the battery in future air transportation networks~\cite{mohsan2022comprehensive}. Therefore, it is more reasonable to consider the battery-charging time cost in the routing problem for electric aircraft.

On the other hand, existing studies of aircraft routing problems fail to consider the congestion effect and battery constraints of electric aircraft in UAM. The existing works in traditional aircraft routing have investigated the dynamic routing strategy for an aircraft that minimizes the expected delay resulting from bad weather~\cite{nilim2002robust,nilim2003multi,balaban2017dynamic}, and aircraft routing problem under a dynamic and uncertain environment~\cite{hong2021anytime}. The studies of UAM routing problems are still in an early stage, such as modifying current-day helicopter routes~\cite{verma2019exploration}, and developing an early methodology for UAM network transport simulation~\cite{rothfeld2018agent}. However, only~\cite{rothfeld2018agent} considers the limited landing capacities at the vertiports. Moreover, none of them considers queues before landing or battery recharge, although~\cite{rothfeld2018agent} takes battery recharge as a possible future extension.


We study the aircraft routing problem that accounts for limited battery capacity, uncertain travel times, and limited landing capacities. Our goal is to route an electric aircraft from its origin to its destination with the minimal expected total travel time while avoiding battery exhaustion.
The contributions are as follows. First, we present a model for the UAM networks that consider the limited battery capacity of aircraft, stochastic travel times through corridors, limited landing (or charging) capacities at vertistops, and queues with uncertain lengths at vertistops. We then provide a sufficient condition for the existence of solutions to the aircraft routing problem based on the model. 
Second, we convert the aircraft routing problem in UAM to a Markov decision process (MDP) and show that the optimal policy of the MDP is the solution to our aircraft routing problem with the minimal expected total travel time.
Finally, we demonstrate our approach with a case study on a network with $29$ vertistops and $137$ corridors, and we are able to obtain the solution to the dynamic routing problem within $347$ seconds.

The current work extends the UAM network model in~\cite{wei2021scheduling} that considered uncertainty in travel times through corridors with lower and upper bounds and limited landing capacity at vertistops, but only allowed aircraft to travel through a set of designated routes. The network model in this paper differs greatly from that in~\cite{wei2021scheduling}: we allow the aircraft to choose any possible route consisting of available corridors and model the travel times stochastically. Further, we add a queue at each vertistop that allows the aircraft to hover and wait at the vertistop when landing spots are fully occupied, and add the constraint on the battery capacity of the aircraft.


Our work is the first step of congestion-aware electric aircraft routing in UAM. The results are particularly useful for routing a single flight, but they also have potential applications in routing a competitive or cooperative fleet of aircraft in the UAM network.

\textbf{Notation:} We use $\mathbb{N}_0$ (resp., $\mathbb{N}$) to represent the set of natural numbers with (resp., without) zero, $\mathbb{R}$ (resp., $\mathbb{R}_+$) to represent the set of all (resp., positive) real numbers, $|\cdot|$ to represent the absolute value, and $\lfloor \cdot \rfloor$ to represent the floor function.

\section{UAM Network Model}

\label{sec:network}
We model an urban air mobility (UAM) network with a directed graph $\mc G = (\mc V,\mc E)$, where $\mc V$ is the set of nodes and $\mc E$ is the set of links for the network: nodes are physical landing sites for the aircraft, sometimes called vertistops or vertiports; links are corridors of airspace connecting nodes. Each node $v\in \mc V$ has capacity---the number of landing spots that allows at most one aircraft to stay and charge at any time---$C_v \in \mathbb{N}_0$. We denote the vector of capacities for all nodes $C = \{C_v\}_{v\in \mc V}$. We define $\tau \defineas \mc E \to \mc V$ and $\sigma \defineas \mc E \to \mc V$ so that for all $e = (v_1,v_2) \in \mc E$ where $v_1,v_2 \in \mc V$, $\tau(e) = v_1$ is the tail of $e$ and $\sigma(e) = v_2$ is the head of $e$. Let $S \subseteq \mc V$ (resp., $T \subseteq \mc V$) be the set of nodes that are not the head (resp., tail) of any edge, $S = \{v \in \mc V \mid \sigma(e) \neq v \quad \forall e \in \mc E\}$ and $T = \{v \in \mc V \mid \tau(e) \neq v \quad \forall e \in \mc E\}$. We assume $S \cap T = \emptyset$. A route $\hat{R}$ is a sequence of connected links, 
with $V(\hat{R})$ containing all nodes it travels through.


We then fix $\Delta T \in \mathbb{R}_+$ as a \emph{discrete parameter} in the network model.

Since in reality, the travel time depends on external factors such as weather conditions or a vehicle’s operational capability, we assume that the travel time $T_e$ for each link $e$ is not exact, but follows a probability mass function $f_e: \{\ul{x}_e,\ul{x}_e+\Delta T,\hdots, \ul{x}_e+\hat{k}_e\Delta T \} \to [0,1]$, i.e.,  $f_e(T_e)$ denotes the probability that a realization of travel time through link $e$ is $T_e$ units of time, and $T_e \in [\ul{x}_e, \ol{x}_e]$ where $\ol{x}_e \defineas \ul{x}_e+\hat{k}_e\Delta T$ and we assume $\ul{x}_e/\Delta T \in \mathbb{N}$. We denote the corresponding cumulative distribution function as $F_e(\cdot)$.
We then define the index set of travel time as $K_e = \{1,2, \hdots, \hat{k}_e\}$. As a result, a realization of travel time $T_e$ of link $e$ satisfies $T_e = k \Delta T+\ul{x}_e$ for some $k\in K_e \cup \{0\}$, and the probability that $T_e = k\Delta T +\ul{x}_e $ given $T_e > (k-1)\Delta T +\ul{x}_e$, denoted as $p^e_{k}$ for all $e \in \mc E$, is 
\begin{align}
    \label{eq:p_realization}
    p^e_{k} = \begin{cases}
    \frac{F_e(k\Delta T +\ul{x}_e)-F_e((k-1)\Delta T +\ul{x}_e)}{1-F_e((k-1)\Delta T +\ul{x}_e)} \quad\forall k\in K_e\\ 
    f_e(\ul{x}_e) \quad  k =0 \,.
    \end{cases}
\end{align}
We denote the vector of all lower and upper travel time limits for all links as $\ul{x}$ and $\ol{x}$.


We assume that aircraft are equipped with batteries that follow the Assumption \ref{assum:battery} below. We use the term \emph{battery level} to infer the remaining time that an aircraft is able to travel with its battery.
\begin{assum}[Aircraft Battery]\
\label{assum:battery}
\begin{enumerate}
    \item An aircraft with a fully charged battery can travel or hover for $B = \hat{k}_b \Delta T$ units of time where $\hat{k}_b \in \mathbb{N}$, i.e., the battery level of a fully charged battery is $B$.
    \item An aircraft can only charge its battery to a designated level in the set $S_C = \{0, \Delta B, \hdots, \hat{k}_c \Delta B\}$, where $\Delta B = k_{\Delta B} \Delta T$ and $\hat{k}_c = B/ \Delta B$ for some $k_{\Delta B}\in \mathbb{N}$ such that $\hat{k}_c \in \mathbb{N}$.
    \item An aircraft needs to charge for $\Delta T_c = k_1 \Delta T \leq B$ units of time to increase its battery level from $k \Delta B$ to $(k+1) \Delta B$ for all $k=1,\hdots, \hat{k}_c$, where $\Delta T_c /\Delta T \in \mathbb{N}$, where $k_1 \in \mathbb{N}$.
    \item If an aircraft hovers or flies for $k\Delta T$ units of time, its battery level will decrease by $k\Delta T$ for all $k\in \mathbb{N}_0$.
    \item An aircraft can only charge its battery when it is staying at a landing spot of a vertistop.
\end{enumerate}
\end{assum}

Based on the assumptions on discrete realizations of travel times and Assumption~\ref{assum:battery}, we can divide the battery consumption equally into $\hat{k}_b+1$, so that $S_B = \{0, \Delta T, 2\Delta T, \cdots, \hat{k}_b \Delta T\}$ is the set of all possible battery levels of an aircraft during a trip. We define the index set of battery levels as $K_B = \{0, 1,2, \hdots, \hat{k}_b\}$ and the index set of battery charging levels as $K_C = \{0, 1, 2, \hdots, \hat{k}_c\}$. Notice that $S_C \subseteq S_B$ and that it takes at most $\hat{T}_B \defineas \hat{k}_c \Delta T_c $ units of time to fully charge a battery.


Once an aircraft has arrived at a node $v \in \mc V$, it can choose to land at the node and block a landing spot for a chosen amount of time for charging or not. An aircraft that decides to land at a node $v$ will enter a queue to hover and wait for an available landing spot. The \emph{queue length} $q_v$---representing the number of the other aircraft at $v$ that will start charging no later then the aircraft---follows a probability mass function $p_v(q_v) \defineas [\ul{q}_v,\ol{q}_v]\cap \mathbb{Q} \to [0, 1]$. We assume that the minimal and maximal queue length $\ul{q}_v, \ol{q}_v \in \mathbb{N}_0$, and $\ol{q}_v \geq \ul{q}_v$. 
We denote the vector of queues at all nodes $Q = \{Q_v\}_{v\in \mc V}$. The aircraft in the queue will land at the landing spots in the order of their position in the queue, and we assume the charging time for the other aircraft at the node follow the assumption below.
\begin{assum}[Battery Charging]\
\label{assum:charge}
Once an aircraft arrives at a node, any aircraft in the queue at the node will charge its battery and occupy a landing spot for another $k \Delta T_c $ units of time with probability $P^c_k$ where $k \in K_C$ and $\sum_{k\in K_C, k\neq 0} P^c_k$ = 1.
\end{assum}

On the other hand, an aircraft that decides not to land at $v$ will choose a link $e \in \mc E_+(v)$ and travel along $e$ without charging its battery,  where $\mc E_+(v) = \{e \in \mc E \mid \tau(e) = v\}$ is the set of \emph{connected links} from node $v$.


\begin{definition}[UAM Network]
A \emph{UAM network} $\mc N$ is a tuple $\mc N = (\mc G, C, Q, \ul{x}, \ol{x})$ where $\mc G, C, Q, \ul{x}, \ol{x}$ are the network graph, node capacities, queue lengths, minimum and maximum link travel times as defined above.
\end{definition}

A \emph{demand} of flight in a network $\mc N$ 
defined as above is a pair $(v_{ori},v_{dest})$, where $v_{ori}$,$v_{dest} \in \mc V$ represent the origin and the destination of the demand. We assume that $v_{ori} \in S$, $v_{dest} \in T$, and $v_{ori} \neq v_{dest}$; we also assume that a flight always departs from its origin with fully charged battery. 

We formalize the aircraft actions using \emph{flight strategies}: a flight strategy is an instruction that assigns the next action to the aircraft at every moment. 
A flight following \emph{safe strategies} will travel through a set of connected corridors, charge the battery at chosen intermediate vertistops, and arrive at its destination without battery exhaustion.
We then define our goal as below.

\textbf{Dynamic Routing Problem for Electric Aircraft}: given a demand of flight in a UAM network defined as above, find the {safe strategies} for a flight with the minimal expected total travel time.

We define the \emph{safe route} before giving a sufficient condition for the safe strategies of a flight to exist.

\begin{definition} [Safe Route]
\label{def:safe_route}
A \emph{safe route} in the UAM network $\mc N = (\mc G, C, Q, \ul{x}, \ol{x})$ is a route $\hat{R} = \{e_j\}_{j=1}^{k_{\hat{R}}}$ that satisfies 
\begin{equation}
\label{eq:R_feasible}
    \ol{x}_{e_j}+\hat{T}_B \cdot {\Big\lfloor\frac{\ol{q}_{\sigma(e_j)}}{C_{\sigma(e_j)}}  \Big\rfloor} \leq B,
\end{equation}
for all $j =1 ,\hdots, k_{\hat{R}}$.
\end{definition}


The following theorem provides a sufficient condition for the existence of safe strategies of a flight.
\begin{thm}
\label{thm:chargin_exists}
Given a flight demand $(v_{ori}, v_{dest})$ in a UAM network $\mc N = (\mc G, C, Q, \ul{x}, \ol{x})$, where $\mc G = (\mc V, \mc E)$ and $v_{ori}, v_{dest} \in \mc V$. If there exists a safe route $\hat{R} = \{e_j\in \mc E\}_{j=1}^{k_R}$ such that $\tau(e_{1}) = v_{ori}$ and $\sigma(e_{k_R}) = v_{dest}$, then we can find the safe strategies for the flight.
\end{thm}


\begin{pf}
To prove Theorem~\ref{thm:chargin_exists}, we first notice that the waiting time for a flight in the queue at a node $v\in V$ is less than $\hat{T}_B \cdot {\Big\lfloor \ol{q}_{v} / C_{v}  \Big\rfloor}$. Therefore, if an aircraft departing from the origin travels through the links along the safe route $\hat{R}$ as defined in the theorem statement, stop and fully charge its battery at each intermediate stop along the route, it can then arrive at the destination without battery exhaustion according to Definition~\ref{def:safe_route}. Hence the safe strategies exist for the flight.
\qed\end{pf}

\section{Markov Decision Process For Dynamic Routing}
\label{sec:MDP_model}

We convert the dynamic routing problem for electric aircraft to a Markov decision process (MDP).

An MDP is a tuple $\mc M = (S, A, P, \mc R, s_0)$, where $S$ is a finite set of states, $A$ is a finite set of actions, $P \defineas A \times S \times S\to [0,1]$ is a state transition probability function, $\mc R \defineas A \times S \times S\to \mathbb{R}$ is a reward function, and $s_0 \in S$ is the initial state. We define the set of \emph{available actions} as $A_s \defineas \{a \in A \mid \exists s' \in S \text{ s.t. }  P(s,a,s') \text{is defined}\}$.
A \emph{path} through $\mc M$---induced by a policy function $\pi \defineas S \to A_s$---is a sequence $R^{\pi} = s_0 a_0 s_1 a_1 \hdots$, where $a_i = \pi(s_i)$ and $P(s_i,a_i, s_{i+1})> 0$ for all $i \geq 0$. 
The objective of an MDP is to find a policy function $\pi$ that maximizes the \emph{discounted expected reward} of the paths induced by $\pi$, $\mathbb{E} \Big[\sum_{i=0}^{\infty} \gamma^i R(s_i,\pi(s_i),s_{i+1})\Big]$, and we denote this policy as the \emph{optimal policy} for the MDP.

We assume that a flight is departing from origin $v_{ori}$ at time $t=0$, and define the states, actions, state transition function, and reward function of the MDP so that at any time $t = k\Delta T$, the status of the flight---including the position information, link-travel time, queue-waiting time, and battery level of the aircraft---corresponds to a state in $S$, an action of the flight corresponds to an action in $A$, the transition of aircraft status follows the transition function $P$, and the travel-time cost corresponds to the reward function. We then demonstrate that the optimal policy for the MDP is the solution to the dynamic routing problem for electric aircraft.

\subsection{States}
\label{sec:states}
Each state of the MDP is given by $s = (v,e,k_e,k_b)$, with $v \in \mc V$, $e\in \mc E$, $k_e \in K_e$, and $k_b \in K_B$. 
To interpret, $v,e$ indicates the related node and edge of the state, $k_e$ implies the realized travel time through the link $e$, and $k_b$ shows the battery level, if applicable. We then divide the states into four cases and explain their relations with flight status in detail.

\underline{\textbf{Case 1:}} The state $s = (v,0,0,k_b)$ represents a status when the aircraft is in the queue at the node $v\in \mc V \notin S$ with battery level $k_b \Delta T$, where $k_b \in K_B\backslash \hat{k}_b$. 
We define the set 
\begin{equation}
\label{eq:S_qv}
    S^q_v = \{s = (v,0,0,k_b) \mid k_b \in K_B\backslash \hat{k}_b\} \,
\end{equation}
for all $v \in \mc V \backslash S$ and let $S^q = \cup_{v \in \mc V\backslash S} S^q_v$. 

\underline{\textbf{Case 2:}} The state $s = (v,-1,-1,k_b)$ indicates a status when the aircraft is charging at the node $v\in \mc V \notin S$ with battery level $k_b \Delta T \in S_B$. An aircraft may start charging at any battery level $b \in S_B$ and stop charging at a battery level $b' \in S_C$ such that $b'>b$.
We designate this set of battery charging states at node $v$ as $S^b_v$, so that 
\begin{equation}
\label{eq:S_bv}
    S^b_v = \{s = (v,-1,-1,k_b) \mid \forall  k_b \in K_B \} 
\end{equation}
for all $v\in \mc V\backslash S$. Because we assume that the aircraft always depart from a node in $S$ with fully charged battery, we let $S^b_v = \{ (v,-1,-1,\hat{k}_b)\}$ for all $v \in S$ and let $S^b = \cup_{v\in \mc V} S^b_v $.

\underline{\textbf{Case 3:}} The state $s = (\tau(e),e,k_e,k_b)$ represents a status when the aircraft travels through a link $e \in \mc E$ with battery level $k_b \Delta T \in S_B$ and realized travel time $\ul{x}_e+k_e\Delta T$.
We let the set 
\begin{equation}
\label{eq:S_le}
    S^l_e = \{s = (\tau(e),e,k_e,k_b) \mid \forall  k_e \in K_E, k_b \in K_B\} 
\end{equation}
denote the set of all link-traveling states through a link $e \in \mc E$, and let $S^l = \cup_{e\in \mc E} S^l_e$.

\underline{\textbf{Case 4:}} The state $s = (\sigma(e), e, 0, k_b)$ demonstrates a status when the aircraft has just arrived at the head node of a link $e\in \mc E$ through the link $e$ with battery level $k_b \Delta T \in S_B$. We denote them as decision-making states, because the aircraft needs to decide the next step: land at the node $\sigma(e)$ and charge, or leave the node and continue to travel on a chosen connected link $e' \in \mc E(\sigma(e))$.
For any $e \in \mc E$, the set of decision-making states is denoted as $S^d_e$, and 
\begin{equation}
\label{eq:S_de}
    S^d_e = \{s = (\sigma(e),e,0,k_b) \mid \forall  k_b \in K_B\} \,.
\end{equation}
Again, $S^d = \cup_{e\in \mc E} S^d_e$. 

Beside the four cases, we add a \emph{target} state $s_t$ that indicates the end of a journey
Therefore, the set of all states, $S$, can be represented as
\begin{equation}
    S = S^q \cup S^b \cup S^l \cup S^d \cup \{s_t\}\,.
\end{equation}
As we assume the flight will depart from the origin $v_{ori}$ with fully charged battery, the initial state of the MDP will be $s_0 = (v_{ori},-1,-1,\hat{k}_b)$.

\subsection{Actions}
\label{sec:actions}

We let the action $a_e$ to represent choosing link $e \in \mc E$ as the next link to travel through, the action $a_c$ to indicate battery charging, and the action $a_0$ to imply a ``default" action that the UAV does not make any new decision.
We define the actions space 
\begin{equation}
    A = \{a_e \mid \forall e\in \mc E\} \cup a_c \cup a_0 \,.
\end{equation}

We then define the set of available actions $A_s$ for all $s\in S$.
\underline{\textbf{Case 1:}} Suppose $s \in S_v^q$, then: $A_s = \{a_c\}$ for all $v\in \mc V\backslash S$ in order to land and charge. 

\underline{\textbf{Case 2:}} Suppose $s = (v,-1,-1,k_b)\in S^b_v$. According to our battery-charging assumption, for all $v \in \mc V\backslash v_{dest}$: if $k_b \Delta T \in S_B\backslash S_C$, the aircraft must continue charging, thus $A_s = \{a_c\}$; 
if $k_b \Delta T \in S_C\backslash B$, the aircraft can either leave for the next link or continue charging, thus $A_s = \{a_e \mid e\in \mc E_+(v)\}\cup\{a_c\}$; 
if $k_b \Delta T =  B$ ($k_b = \hat{k}_b$), the aircraft will leave for the next link, thus $A_s = \{a_e \mid e\in \mc E_+(v)\}$.
But when $v = v_{dest}$, the aircraft does not need further action for arrival, and thus $A_s = \{a_0\}$.

\underline{\textbf{Case 3:}} Suppose $s \in S_e^l$. The aircraft will continue traveling along the link $e$ by default, then $A_s = \{a_0\}$ for all $e \in \mc E$.

\underline{\textbf{Case 4:}} Suppose $s \in S_e^d$, then the aircraft can choose to land at the node $\sigma(e)$ and charge, or to leave this node and travel through the chosen link $e' \in \mc E_+(\sigma(e))$. Hence, $A_s = \{a_e \mid e\in \mc E_+(\sigma(e))\}\cup\{a_c\}$.

\subsection{State Transition Function and Reward Function} 
\label{sec:P_R}

We then construct a state transition probability function $P \defineas A \times S \times S \to [0,1]$ indicating the aircraft execution, and a reward function $R \defineas A \times S \times S \to \mathbb{R}$ indicating the travel time cost, the out-of-battery penalty, and the arrival reward. 
We define several basic reward multiplier for easier representation:
$r_t <0 $ represents the travel-time cost for every $\Delta T$ units of time, $r_b \defineas r_t\cdot \hat{T}_B/B$ indicates the travel time cost to refill the battery level by $\Delta T$, $r_d<0$ is the battery-exhaustion penalty, and $r_a >0$ is the arrival reward.



\underline{\textbf{Case 1:}} Suppose $s =(v,0,0,k_b) \in S^q_v$, then taking the only action $a_c$ will lead to either a charging state $s^b_{k_b-k} = (v,-1,-1,k_b-k) \in S^b_v$ with the probability $p^q(k \Delta T)$---which will be computed next---for some $k =0,\dots, k_b-1$ after waiting in the queue for $k \Delta T$ units time, or the zero-battery queue state $(v,0,0,0) \in S^q_v$ with the probability $p^q_0 = 1- \sum_{k = 0}^{k_b} p^q(k \Delta T)$. 


Let $q_v \in \mathbb{N}_0$ be the queue length at node $v$, so that $q_v \in [\ul{q}_v, \ol{q}_v]$. When $q_v \geq C_v$, according to Assumption~\ref{assum:charge}, we then compute the probability that the $j$'th aircraft in the queue at node $v$ will charge for $k^c_j \Delta T_c$ units of time, where $k^c_j \in K_C\backslash \{0\}$ for all $j\in 1,2, \dots, q_v$, as below:
\begin{equation}
\label{eq:p_qlist}
    P^{\pi}_{(k^c_1,k^c_2, \dots,k^c_{q_v})} = \prod_{j = 1}^{q_v} P^c_{k^c_j}\,,
\end{equation}
We then compute $p^q(k\Delta T)$ with Algorithm~\ref{alg:queue_time} and define



\begin{align}
  \label{eq:P_queue}
  P(s,a_c,s') &= \begin{cases}
                  p^q(k \Delta T) & \text{if } s' =s^b_{k_b-k}, k_b > k \\
                  p^q_0 & \text{if } s' = (v,0,0,0), k_b \leq k \,.
\end{cases}
\end{align}
and
\begin{align}
  \label{eq:R_queue}
  R(s, a_c, s') &= \begin{cases}
                 k \cdot r_t & \text{if } s' =s^b_{k_b-k}, k_b > k \\
                  r_d & \text{if }  s' = (v,0,0,0), k_b \leq k \,.
\end{cases}
\end{align}

\begin{algorithm}[t]
\caption{Event-triggered Scheduling Algorithm}
\label{alg:queue_time}
\begin{algorithmic}[1]
\Statex \textbf{Inputs:} {$v\in \mc V$, $\ul{q}_v$, $\ol{q}_v$, $p_v$, $C_v$, $\Delta T_c$, $k \Delta T$}
\Statex \textbf{Outputs:} $p^q(k \Delta T)$
\For{$q = \ul{q}_v, \ul{q}_v+1,\hdots,  \ol{q}_v $}
\If{$q < C_v$}  \Comment{No wait before landing}
\State $p^1_q = 1$ if $k \Delta T = 0$, and $p^1_q = 0$ otherwise
\Else
\State $S_{perm} \defineas \{\vec{k}^c = (k^c_1,k^c_2, \dots,k^c_{q}) \mid k^c_j \in K_C\backslash \{0\}\}$ 
\State $p^1_q =0$
\For{$\vec{k}^c = (k^c_1,k^c_2, \dots,k^c_{q})  \in S_{perm}$} 
\State $w_n = 0$ for $n = 1, \hdots, C_v$
\State $w \defineas (w_1,w_2, \hdots, w_{C_v})$
\For{$j = 1 \to q$} 
\State $n' = argmin_{n' = 1, \hdots, C_v} \{w_1,w_2, \hdots, w_{C_v}\} $ 
\State $w_{n'} = k^c_j \Delta T_c + w_{n'}$, and update $w$ 
\EndFor
\State $t_e = \min\{w_1,w_2, \hdots, w_{C_v}\}$
\If{$t_e == k \Delta T$}
\State Compute $P^{\pi}_{\hat{k}^c}$ according to \eqref{eq:p_qlist}
\State $p^1_q = p^1_q + P^{\pi}_{\hat{k}^c}$
\EndIf
\EndFor
\EndIf
\EndFor
\State $p^q(k \Delta T) = \sum_{q = \ul{q}_v}^{\ol{q}_v} p_v(q) p^1_q$
\end{algorithmic}
\end{algorithm}

\underline{\textbf{Case 2:}} Suppose $s = (v,-1,-1,k_b) \in S^b_v$. Assume $v \neq v_{dest}$, if $k_b \Delta T \in S_B \backslash B$, taking the action $a_c \in A_s$ will lead to the next battery charging state $s_b = (v,-1,-1,k_b+1)$; if $k_b \Delta T \in S_C$, taking the action $a_{e'}\in A_s$ will lead to the link traveling state $s_{e',k_b'} = (\sigma(e), e', 0, k_b') \in S^l_{e'}$, where $k_b' = \max \{k_b - \ul{x}_{e'}/\Delta T,0\}$. 
Otherwise if $v = v_{dest}$, then taking the default action $a_0$ will lead to the target state $s_t$.
We set 
\begin{align}
  \label{eq:P_battery}
  \nonumber & P(s,a,s') = \\ 
  & \begin{cases}
                  1 & \text{if } a = a_c, s' = s_{b}, k_b \Delta T \in S_B \backslash B, v \neq v_{dest}\\
                  1 & \text{if }  a = a_{e'}\in A_s, k_b\Delta T \in S_C, s' = s_{e'}, v \neq v_{dest} \\
                  1 & \text{if }  a = a_0, v= v_{dest}, s'=s_t  \,.
\end{cases}
\end{align}
We let and $\Delta k = k_b - k_b'$, and set
\begin{align}
  \label{eq:R_battery}
  \nonumber & R(s,a,s') = \\
  & \begin{cases}
                  r_b & \text{if } a = a_c, s' = s_{b}, k_b \Delta T \in S_B \backslash B, v \neq v_{dest}\\
                  r_t \Delta k  & \text{if }  a = a_{e'}\in A_s, k_b\Delta T \in S_C, s' = s_{e'}, v \neq v_{dest} \\
                  r_a & \text{if }  a = a_0, v= v_{dest}, s'=s_t  \,.
\end{cases}
\end{align}

\underline{\textbf{Case 3:}} 
Suppose $s =(\tau(e),e,k_e,k_b)  \in S^l$, when taking the only available action $a_0 \in A_s$, if $k_b > 0$ the next state will be either the queue state of equal battery $s_2 = (\sigma(e),e,0,k_b) \in S^q_{\tau(e)}$ with probability $p^e_k$ computed through~\eqref{eq:p_realization}, or the next link-traveling state, $s_3 = (\tau(e),e,k_e+1,k_b-1) \in S^l_e$ with probability $1 - p^e_k$; if $k_b = 0$, the next state will be $s$---indicating battery dies.
We then define 
\begin{align}
  \label{eq:P_link}
  P(s,a_0,s') &= \begin{cases}
                  p^e_k &\text{if } k_b >0, s'=s_2\\
                  1- p^e_k &\text{if } k_b >0,s' = s_3, k_e<\hat{k}_e\\
                  1 &\text{if } k_b =0, s' = s \, ,
\end{cases}
\end{align}
and
\begin{align}
  \label{eq:R_link}
  R(s,a_0,s') &= \begin{cases}
                  0 &\text{if } k_b >0, s'=s_2\\
                  r_t &\text{if }  k_b >0,s' = s_3, k_e<\hat{k}_e\\
                  r_d & \text{if }  k_b =0, s' = s \, .
\end{cases}
\end{align}

\underline{\textbf{Case 4:}} Suppose $s =(\sigma(e),e,0,k_b) \in S^d$. 
The aircraft will enter the queue state $s_q = (\sigma(e),0,0,k_b) \in S^q_{\sigma(e)}$ with the action $a_c$, or the link-traveling state $s_{e',k_b'} = (\sigma(e), e', 0, k_b') \in S^l_{\sigma(e')}$ with the action $a_{e'}\in A_s$, where $k_b' = \max \{k_b - \ul{x}_{e'}/\Delta T,0\}$. 
Therefore, we define
\begin{align}
  \label{eq:P_decision}
  P(s,a,s') &= \begin{cases}
                  1 &\text{if } a = a_c, s' = s_q\\
                  1 & \text{if } a = a_{e'}\in A_s, s' = s_{e',k_b'} \,,
\end{cases}
\end{align}
We let and $\Delta k = k_b - k_b'$, and set
\begin{align}
  \label{eq:R_decision}
  R(s,a,s') &= \begin{cases}
                  0  &\text{if } a = a_c, s' = s_q\\
                  r_t \Delta k & \text{if } a = a_{e'}\in A_s,  s' = s_{e',k_b'} \,.
\end{cases}
\end{align}


Additionally, we set $P(s_t,a,s_t) = 1$ for all $a\in A$ indicating the end of the journey once reaching the target state $s_t$ and set $R(s_t,a,s_t) = 0$ correspondingly.

\subsection{Solving the Dynamic Routing Problem for Electric Aircraft with MDP}


Given a demand of flight $(v_{ori},v_{dest})$, and a UAM network $\mc N$ with a safe route connecting from $v_{ori}$ to $v_{dest}$, we convert the problem of finding safe strategies for the flight into the MDP defined as in Section~\ref{sec:states}--\ref{sec:P_R}. 

We then demonstrate that an optimal policy of the MDP provides us the safe strategies for the flight with minimal travel time.
The initial state of the MDP is $s_0 = (v_{ori},-1,-1,\hat{k}_b)$. 
According to Theorem~\ref{thm:chargin_exists}, 
we can then find a policy $\pi_1$ so that any path $R^{\pi_1} = s_0 a_0 \hdots$ where $a_i = \pi_1(s_i)$ and $s_0 = (v_{ori},-1,-1,\hat{k}_b)$ will have $s_l = s_t$ for some $l \in \mathbb{N}$. 
Our design of reward $R$ ensures that any policy that has nonzero probability to fail to reach the target state or to exhaust its battery will have smaller expected reward than that with $\pi_1$. Moreover, the reward design reflects the travel time cost of the flight.
Therefore, the optimal policy $\pi^*$ of the MDP provides the safe strategies with minimal travel time of the flight. We can find this optimal policy with known algorithms, e.g., value iteration algorithm, policy iteration algorithm, and linear programming.

Therefore, we are able to solve our dynamic routing problem for electric aircraft by finding the optimal policy of the MDP. 


\begin{remark}
Although the network model we study considers the travel times, battery levels, and battery charging times as discrete parameters, we can apply the results to a UAM network with continuous parameters by discretizing the continuous parameters. Moreover, Theorem~\ref{thm:chargin_exists} is also applicable to a UAM network with continuous parameters.
\end{remark}




\section{Case Study}
\label{sec:case_study}

We demonstrate our approach for finding safe strategies for a flight with minimal travel time on a case study.
We consider a UAM network with $29$ nodes and $137$ links inspired by the map around the cities of Austin and Dallas, Texas, and find the optimal policy of the corresponding MDP using value iteration algorithm.

We considered a possible UAM network between Dallas and Austin in Texas with map shown in Fig.~\ref{fig:network_pic}. We took the cities on the map as the set of nodes $\mc V$ of the network graph, and all pairs of cities---with intermediate distance less than 60 miles---as the set of links $\mc E$.
We measured the distance of each link in Google Earth and the resulted network has $29$ nodes and $137$ links\footnote{The details of the map data, case-study settings and Matlab simulation code are in \url{https://github.com/QinshuangCoolWei/Dynamic-Routing-in-Stochastic-UAM-Network.git}}. We show each node in Fig.~\ref{fig:network_pic}, but decided not to label any link due to the large quantity.

We assumed the following: the battery capacity is $B = 30$ minutes, the top speed of the aircraft is $v_{max} = 150$ mph, the time to fully charge a battery from zero is $\hat{T}_B  = 6$ minutes, and the maximal battery indicator is $\hat{k}_b = 2$, i.e., the aircraft can charge its battery level to $15$ or $30$ minutes.
We then construct the UAM network model with $\Delta T$ by approximating the lower and upper travel time of a link $e$ as below to satisfy the assumptions for the model:
\begin{align*}
    \ul{x}_e & = round \Big\{\frac{dist(e)}{v_{max} \cdot\Delta T} \Big\}\cdot \Delta T \\
    \ol{x}_e & =\Big \lfloor  \frac{\ul{x}_e \cdot 1.2 }{\Delta T}\Big\rfloor   \cdot \Delta T+\Delta T\,, 
\end{align*}
where $dist(e)$ is the measured distance between the two nodes (or cities). We set the travel-time cost $r_t = -\Delta T$, the battery-exhaustion penalty $r_d = -1000$, and the arrival reward $r_a = 1000$.

First, we considered a flight traveling from Dallas to Austin, and demonstrated the solutions to our dynamic routing problem for the flight under two different network settings. We made up the capacities and two sets of deterministic queue lengths for the nodes, and denoted one set as the \emph{origin} setting while the other as the \emph{comparison} setting. We show the capacities and queue lengths of all nodes under the two settings in Fig.~\ref{fig:network_pic}.

For both $\Delta T = 1$ and $\Delta T = 3$, the optimal policies of the corresponding MDPs (the solutions to the dynamic routing problem of the flight) computed by the value iteration algorithm are the same, regardless the origin or the comparison setting. In both settings, the MDPs have $11952$ states with $\Delta T = 1$ and we found the solutions in $18818$ seconds; they have $3414$ states with $\Delta T = 3$ and we found the solutions in $347$ seconds. Fig.~\ref{fig:network_pic} demonstrates two routes induced by the two solutions: in the origin setting, the aircraft will land at each intermediate stop and fully charge its battery; in the comparison setting, the aircraft will land at each intermediate stop except the Pflugerville node and fully charge its battery, but will only need to land and charge its battery if the battery level is less then $15$ minutes upon its arrival at the Pflugerville node.

The difference between the two routes in Fig.~\ref{fig:network_pic}---two possible routes induced by the two solutions for the origin and the comparison settings---demonstrates the strategy adjustments due to the possible waiting time at the nodes. With the capacity unchanged at a node, if the queue length exceeds the capacity, a longer queue length leads to a longer waiting time before landing. Therefore, when we convert the network from the origin to the comparison setting, the longer queue lengths at the Gatesville and the Georgetown nodes drive the flight away, and the shorter queue lengths at the Pflugerville and the Round Rock nodes attract the flight to them. Fig.~\ref{fig:network_pic} conforms to the finding above: the route with the comparison setting (blue, dashed line) gets rid of the Gatesville and the Georgetown nodes, which appear in the route with the origin setting (purple, solid line), and lands at the Pflugerville node. 

Moreover, since the longest travel time between the Pflugerville node and the destination---the Austin node---is less then $15$ minutes for both $\Delta T = 1$ or $3$, a battery level higher or equal than $15$ minutes will guarantee the arrival at the destination without battery exhaustion from the the Pflugerville node.


\begin{figure}[!t]
\centering
\input{pic/map.tikz}
\caption{The map around two cities in Texas, Austin and Dallas captured from Google Earth. For each node/city, we use the shape to denote the capacity, and the color to denote the queue length. A colored shape beside a node without border denotes the origin setting, and the one with border represents the comparison setting. Correspondingly, the path connected by the solid purple lines and curves shows a route induced by the strategy obtained from the origin setting; the path connected by the dashed blue lines and curves show a route induced by the strategy obtained from the comparison setting. (We use a curve representation only to avoid passing over other nodes; the implied corridor is actually a straight line.) }
\label{fig:network_pic}
\end{figure}

Second, we demonstrated the importance of taking the queues into consideration in UAM networks.
For all nodes in the network except Austin, we found the solutions to the dynamic routing problem for the flight from the node to Austin while setting $\ol{q}_v = 0$ for all $v \in  \mc V$. For each origin node $v_0$, we considered the longest travel time through any link chosen by the flight strategies in the solution, and denoted the induced route as the \emph{worst-case optimal route} for the aircraft to travel from $v_0$ to Austin. We generated $28$ worst-case optimal routes for the aircraft to travel from the $28$ origin nodes to the Austin node. 

\begin{figure}[!t]
\centering
%
%
\pgfplotsset{width=7cm,compat=1.16}

\begin{tikzpicture}

\pgfplotsset{set layers}

\begin{axis}[%
width=7 cm,
height=4 cm,
scale only axis,
xmin=-0.1,
xmax=4.1,
axis x line*=bottom,
x axis line style=blue,
xlabel style={font=\color{blue}},
xlabel={$\ol{q}_{\sigma(e)}$},
ymin=0.1,
ymax=1.1,
ylabel={$r_{safe}$},
legend style={at={(axis cs:3.87,1.05)},draw=none}
]

\addplot [color=blue, dotted, mark=o, mark options={solid, blue}]
  table[row sep=crcr]{%
0	1\\
1	0.905109489051095\\
2	0.64963503649635\\
3	0.430656934306569\\
4	0.27007299270073\\
};
\addlegendentry{Link}
\end{axis}

\begin{axis}[
width=7 cm,
height=4 cm,
scale only axis,
xmin=-0.1,
xmax=4.1,
ymin=0.1,
ymax=1.1,
axis x line*=top,
xlabel={$\max_{e\in \hat{R}}\ol{q}_{\sigma(e)}$},
legend style={at={(axis cs:4,0.9)}, draw = none},
]
\addplot [color=black, mark=o, mark options={solid, black}]
  table[row sep=crcr]{%
    0	1\\
    1	0.771084337349398\\
    2	0.521885521885522\\
    3	0.35264483627204\\
    4	0.192045454545455\\
    };
\addlegendentry{Route}
\end{axis}

\end{tikzpicture}%
\caption{The graph for the ratio $r_{safe} = \frac{n_{safe}}{n_{total}}$ for the case of route and link. We use the black, solid (resp., blue, dotted) line to represent the case of route (resp., link). In the link case, the x axis shows the maximal queue length at the head node of the link; in the route case, the x axis shows the largest maximal queue length for all nodes except the origin along the route.
}
\label{fig:q_pic}
\end{figure}
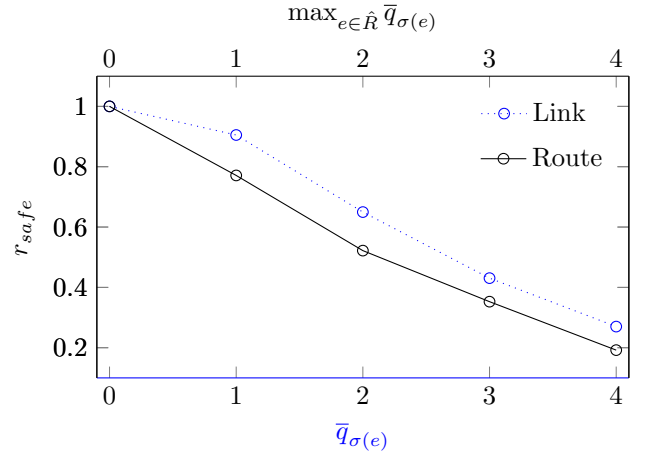

Given an origin node $v_0$, we considered the worst-case optimal route starting from $v_0$. We then tried different combinations of maximal queue lengths from 0 to 4 for all nodes along the route $\hat{R}$, i.e., $\ol{q}_v \in \{0,1,\hdots,4\}$ while $\ul{q}_v =0$ for all $v\in V(\hat{R})\backslash v_0$, and tested if the route is still a safe route. In this case, we define the ratio $r_{safe} = {n_{safe}}/{n_{total}}$, where $n_{safe}$ is the sum of the total number of combinations with which the route is still a safe route across all $28$ routes, and $n_{total} = 3520$ is the sum of the total number of all combinations across all $28$ routes.
Similarly, for all $e \in \mc E$ in the network, we tested if the link still satisfies~\eqref{eq:R_feasible} with $\ol{q}_{\sigma(e)} \in \{0,1,\hdots,4\}$ and $\ul{q}_{\sigma(e)} =0$, that is, the flight with fully charged battery is able to travel through the link and land at its head node without battery exhaustion. In this case, the ratio is still $r_{safe} = {n_{safe}}/{n_{total}}$, but $n_{safe}$ is the sum of the total number of links that satisfy~\eqref{eq:R_feasible}, and $n_{total} = 137$ is the number of all links. The relations between $r_{safe}$ and maximal queue length for both the link and the route cases are plotted in Fig.~\ref{fig:q_pic}, where the maximal queue length stands for $\ol{q}_{\sigma(e)}$ of the link $e$ in the link case, and stands for the largest $\ol{q}_{v}$ among all nodes except the origin along the route in the route case.

In Fig.~\ref{fig:q_pic}, for the route case, the ratio $r_{safe}$ is $1$ when $\ol{q}_{v}=0$ for all $v \in \mc V$, but the ratio decreases when the largest $\ol{q}_{v}$ among all nodes along the route increases. 
When $\ol{q}_{v}=0$ for all $v \in \mc V$, all the links satisfy~\eqref{eq:R_feasible}, hence all the $28$ routes are safe routes. However, when the maximal queue length increases at the head nodes of the links, some links no longer satisfy~\eqref{eq:R_feasible}, as we can observe from the link case in Fig.~\ref{fig:q_pic}, and thus the routes that contain any of these links are no longer safe routes according to Definition~\ref{def:safe_route}.

As a result, if we fail to consider the queues at the nodes in the UAM network, i.e., falsely take $\ol{q}_{v} =0$ for all $v\in \mc V$, the aircraft may exhausts its battery during its flight, as the we may generate the strategies that assign the aircraft to travel through a route that is in fact not a safe route with consideration of queues.

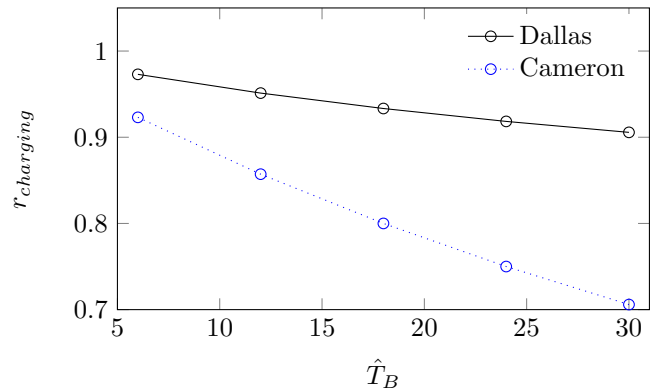
\begin{figure}[!t]
\centering
%
%
\begin{tikzpicture}

\begin{axis}[%
width=7 cm,
height=4 cm,
scale only axis,
xmin=5,
xmax=31,
xlabel={$\hat{T}_B$},
ymin=0.7,
ymax=1.05,
ylabel={$r_{charging}$},
legend style={legend cell align=left, align=left, draw=none}
]
\addplot [color=black, mark=o, mark options={solid, black}]
  table[row sep=crcr]{%
6	0.972972972972973\\
12	0.951219512195122\\
18	0.933333333333333\\
24	0.918367346938776\\
30	0.905660377358491\\
};
\addlegendentry{Dallas}

\addplot [color=blue, dotted, mark=o, mark options={solid, blue}]
  table[row sep=crcr]{%
6	0.923076923076923\\
12	0.857142857142857\\
18	0.8\\
24	0.75\\
30	0.705882352941177\\
};
\addlegendentry{Cameron}

\end{axis}
\end{tikzpicture}%
\caption{The graph for the ratio of $r_{charging} = \frac{t_{naive}}{t_{optimal}}$. The black, solid (resp., blue, dotted) line represents the flight traveling from Dallas (resp., Cameron).}
\label{fig:charge_pic}
\end{figure}

Finally, we observed that even the flight traveling through the worst-case optimal routes can benefit from our charging strategies---the aircraft will not necessarily fully charge its battery at each intermediate node along its route---comparing to the \emph{naive} charging strategy where the aircraft stop and fully charge its battery at each intermediate node along its route. We let the ratio $r_{charging} = {t_{optimal}}/{t_{naive}}$, where $t_{optimal}$ is the shortest total travel time for the flight to travel through the worst-case optimal route with the charging strategies we generated, and $t_{naive}$ is that with the naive charging strategies. We chose the two flights traveling from Dallas and Cameron along the worst-case optimal routes, and plotted the relation between $r_{charging}$ and the maximal charging time $\hat{T}_B$ in Fig.~\ref{fig:charge_pic}.

In Fig.~\ref{fig:charge_pic}, $r_{charging}$ decreases when the maximal charging time increases for both the Dallas and the Cameron routes: $r_{charging}$ decreases significantly by more than 0.2 in the flight of Cameron, and decreases by less than 0.1 in the flight of Dallas. We can infer that, the flights generally save more time from our charging strategies when the maximal charging times are longer, but the benefit varies with different networks and different flights.

\section{Conclusion}

We studied the dynamic routing problem for electric aircraft that looks for safe strategies for a flight with minimal expected total travel time. We provided a model for UAM networks with limited battery capacity of the aircraft, stochastic travel times through corridors, and limited landing capacity and queues at vertistops. We then transfer the routing problem to a Markov decision process, and show that the optimal policy of the MDP is the solution to the dynamic routing problem for electric aircraft in the UAM network. We therefore conclude that given a UAM network, we can find the solution to the dynamic routing problem by finding the optimal policies of the corresponding MDP.

However, the current study only provides strategies for a single aircraft, ignoring the possible interactions of aircraft in the UAM network. We will study the strategies when collaboration and competition of aircraft are taken into account.

\bibliography{ifacconf}             
                                                   







\end{document}